# UMTS-WiMAX Vertical Handover in Next Generation Wireless Networks

Nada Alamri[1] and Nadine Akkari[2]

[1]Department of Computer Science, King Abdulaziz University, Jeddah, KSA
[1]nalamri062@stu.kau.edu.sa
[2]nakkari@kau.edu.sa

## ABSTRACT

*The vision of next generation wireless network (NGWN) is to integrate different wireless access technologies, each with its own characteristics, into a common IP-based core network to provide mobile user with service continuity and seamless roaming. One of the major issues for the converged heterogeneous networks is providing a seamless vertical handover (VHO) with QoS support. In this paper we have reviewed the various interworking architectures and handover scenarios between UMTS and WiMAX. Also, we have compared the proposed solutions based on different criteria and revealed the pros and cons of each scheme. The comparison aids to adopt a better interworking and handover mechanism in NGWN.*

## 1. INTRODUCTION

4G is the short term for fourth-generation wireless, the stage of broadband mobile communications that will supersede the third generation (3G). While neither standards bodies nor carriers have concretely defined or agreed upon what exactly 4G will be, it is expected to provide end-to-end IP services with higher bandwidth and data rate, lower cost and authentication overhead, and with service constantly provided to the user without any disruption. Another feature that 4G is expected to provide is high level of user-level customization. That is, each user can choose the preferred level of quality of service, radio environment, etc.

For the migration towards 4G, the standardization bodies have been working on developing new access networks that promise to increase throughput and ease network load such as: Mobile WiMAX (Worldwide interoperability for Microwave Access), Evolved UTRAN (E-UTRAN) and Ultra Mobile Broadband (UMB). The Mobile WiMAX can be considered as a superior 4G technology designed to provide for 4G services, beyond the current 3G technologies' horizon. The UTRAN, UMB and WiMAX are based on OFDMA plus MIMO that significantly increase data rates for mobile terminals, reduce end-to-end latency for real-time communications and set-up times for new connections [1]. However, the upgrading process to 4G standards is expensive, difficult, and time consuming. Therefore during the transition period, integrating a newer faster network with existing and evolving networks infrastructure seems to be the optimal option for providing user with always on connectivity and allowing service providers to incrementally upgrade their networks while maintaining coverage. Accordingly, integration of heterogeneous networks, known as Next Generation Wireless Networks (NGWN), is receiving a lot of attention in standardization forums such as the 3GPP, 3GPP2, IETF, WiMAX Forum, and IEEE 802 LAN/MAN Committee. In this context, integration of WiMAX with other wireless technologies, such as WLAN and 3G, has attracted research community for the last few years since the Mobile WiMAX is the only access technology of 4G that is available today. However,

WiMAX networks are still in its infancy with many technical challenges lying ahead. The main hindrance to its competitive edge is the lack of an infrastructure for the WiMAX Core Network (CN), which makes the deployment of WiMAX on its own a far future [2]. Besides that it is impossible to become a head-to-head competitor against current cellular networking infrastructures [3]. Therefore, integrating Mobile WiMAX with the already well established 3G cellular networks such as UMTS (Universal Mobile Telecommunications System), will make the best use of advantages of both technologies and eliminate their stand-alone defects [3, 4].

One of the most challenging issues for seamless internetworking of heterogeneous networks is the ability maintain an ongoing session when Mobile Station (MS) changes access networks while maintaining session Quality of Service (QoS). The process of transferring session between heterogeneous networks is known as Vertical Handover (VHO)[5].

In this paper we provide a survey of the existing VHO solutions between UMTS and WiMAX networks concerning integration architecture and mobility management protocols. We present a novel architectural classification for these solutions, review and compare the schemes based on different criteria including, architecture, mobility management and handover performance. The rest of this paper is organized as follows. Section 2 discusses related work in the area of mobility management protocols. Overview of UMTS-WiMAX integration architectures is presented in section 3. Various handover scenarios for UMTS-WiMAX are classified and discussed in section 4. In section 5, we provide a comparison of existing VHO solutions between UMTS and WiMAX. In the last section we conclude and discuss future works.

## 2. RELATED WORK

Mobility management on different layers of the protocol stack may be performed in different ways with different protocols. Mobile IP (MIP) is a network layer protocol proposed by IETF to allow MS to remain reachable in spite of its movements within wireless IP environments [6]. Several MIP enhancements exist today such as Hierarchical MIP (HMIPv6, proposed to reduce global signaling load during intra domain handover), Fast Handovers for MIPv6 (FMIPv6, a way of pre-binding process to reduce the required time to obtain a new IP)[7], and Proxy MIPv6 (PMIPv6, proposed to reduce a signaling delay in confined areas)[8]. On the other hand, Mobile Stream Control Transmission Protocol (mSCTP) is an extension of STCP, an IETF transport layer protocol with multi-homing feature, with Dynamic Address Reconfiguration (DAR) feature. An mSCTP enabled MS has the ability of acquiring and holding multiple IP addresses while keeping the end-to-end connection intact. Also, as an application layer solution, Session Initiation Protocol (SIP) was proposed for establishing, modifying, and terminating multimedia sessions [9]. SIP has high flexibility and provides support for terminal, session, personal and service mobility[10]. Each of these mobility management protocols has features and drawbacks.

Many solutions to improve VHO performance for these protocols have been reported in the literature. [11] and [12] have combined the functionality of transport layer protocols with SIP to allow MS to use multiple IPs. [11] used SCTP with SIP to support real-time services for VHO, taking advantage of multi-homing feature of SCTP bind together with the location and network transparency provided by SIP with no need for the re-invite message signalling. In [12], mSCTP is used with SIP in IMS-based heterogeneous environment where a proxy based on mSCTP is setup between application server and MS. The proxy acts as an anchor point for soft VHO of MS with multiple interfaces. On the other hand, [13], [14] and [15] used Media Independent Handover (MIH) to enhance the performance of vertical handovers in heterogeneous networks for SIP, FMIPv6 and MIP, respectively.

## 3. UMTS-WIMAX INTEGRATION ARCHITECTURES

Before proceeding with the description of VHO solutions for integrated UMTS-WiMAX networks, it is necessary to review the interworking architecture models. The interworking architectures between UMTS and WiMAX networks can be classified into: loose, tight and very tight coupling architectures.

Loose coupling indicates that the interworking point is after the GGSN through Gi interface, where the two networks are connected independently but utilizing a common subscription. It has the advantage of simplicity with minimal enhancements to existing components, but at the expense of considerably larger handover execution time. On the other hand, tight coupling provides interworking at the UMTS CN level where the WiMAX network may emulate a RNC or a SGSN. The WiMAX data and signalling traffic are transferred through the UMTS CN and it uses the same authentication, mobility, and billing infrastructures [16, 17]. Loose coupling is illustrated in Figure 1.

The tight coupling can provide seamless VHO and is the preferred model for the service providers' side due to its possibility to enable control of users who are using WiMAX access. However, tight coupling requires technology-specific modifications on WiMAX AN and UMTS core network as well, where WiMAX AN must implements 3G radio access network protocols to route traffic through the core 3G elements and to communicate with the 3G network and SGSN and GGSN nodes need to be extended to support the large amount of data from WiMAX users through the UMTS network [16]. In the very tight coupling architecture, also known as integrated coupling, the WiMAX is integrated in 3G network on the access network level. The BS of WiMAX is connected directly to the RNC of UMTS. The WiMAX is considered as another radio interface and the integration is achieved below IP, by reusing some UTRAN- level mobility mechanisms. Similarly to tight coupling, many technology-specific modifications are needed, but major modifications take place in the UTRAN, while the CN PS is upgraded to support the additional traffic from the WiMAX. Very tight coupling is expected to provide best VHO performance in exchange of high cost modifications for both AN and CN, but it is only suitable for a single operator for security and management reasons. Tight and very tight coupling are shown in Figure 1.

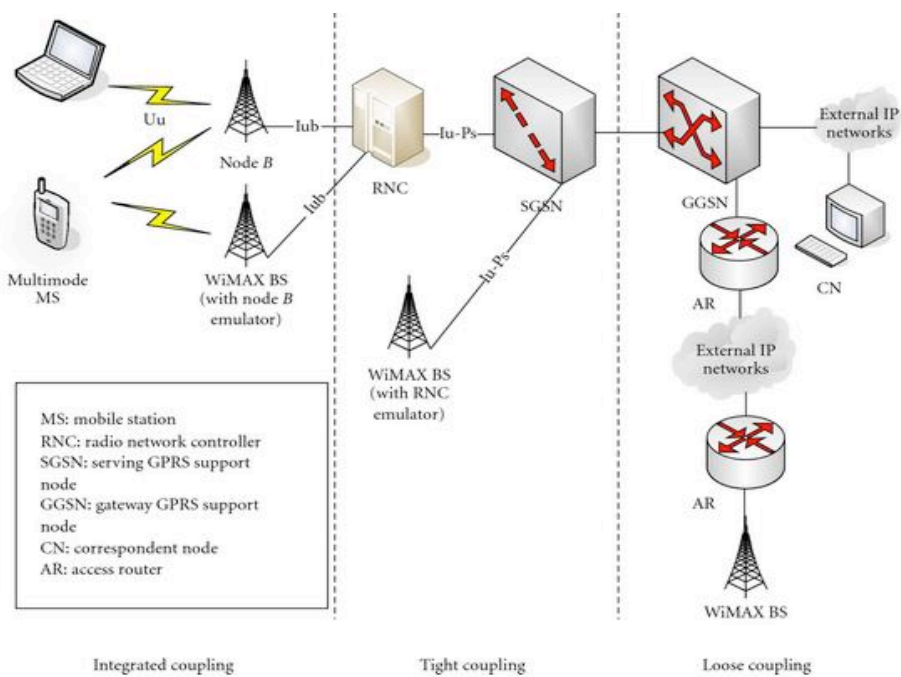

Figure 1. UMTS-WiMAX coupling architecture models

## 4. UMTS-WiMAX Handovers

Currently, there are several research proposals in the literature that discuss the integration of 3G and WiMAX networks. They mainly focus on providing mobile users with service continuity and seamless roaming when switching between the two access networks. Many solutions have been proposed and we have studied them in function of the integration architecture, mobility management protocols and handover (HO) performance (packet loss, QoS and signalling load) and classified them from an architectural point of view into: loose and tight coupling, IMS-based and EPC-based solutions. Details about the classified solutions are described below.

### 4.1. Loose and tight coupling solutions

Initial attempts to integrate UMTS and WiMAX have been made by using different flavours of the loose and tight coupling architectures.

[16] presented a loosely coupled network architecture and MS architecture. The proposed architecture includes procedures for activating a QoS session, mapping between network-specific QoS classes, and network and session layer QoS support. To achieve the interworking of the two networks specific new components are added. Figure 2 shows integration network architecture of UMTS and WiMAX. The FAs of MIP are deployed in WiMAX Gateway and GGSN of UMTS, while HA, common home AAA server and SIP server are deployed in the Internet to be accessible by both networks. Each network has a local AAA, which is interfaced with home AAA to authenticate subscriber. The MS software model contains a mobility and session manager. Mobility manager is responsible for HO decision and initiation and provides mobility with MIP. It also, provides feedback to session manager, which is responsible for translating application requirements into QoS parameters specific to the network and provides feedback for real time application to adjust to the network characteristics and inform CN by using SIP and SDP. Network layer QoS is achieved using DiffServ while session layer QoS is maintained using SIP for real time application. Real-time session HO is shown in Figure 3.

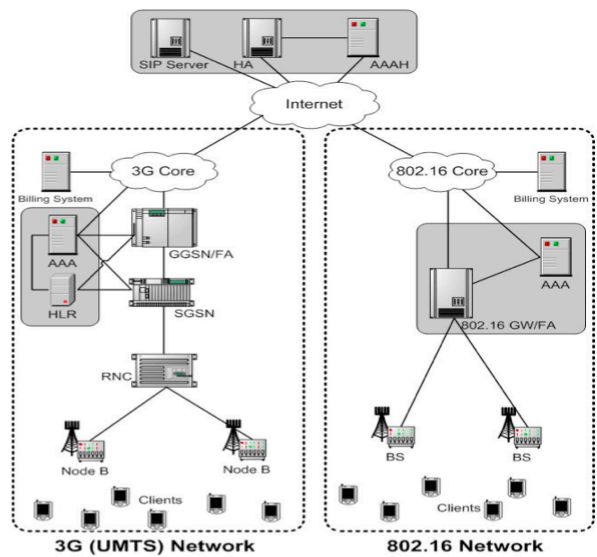

Figure 2. Integrated architecture of UMTS and WiMAX

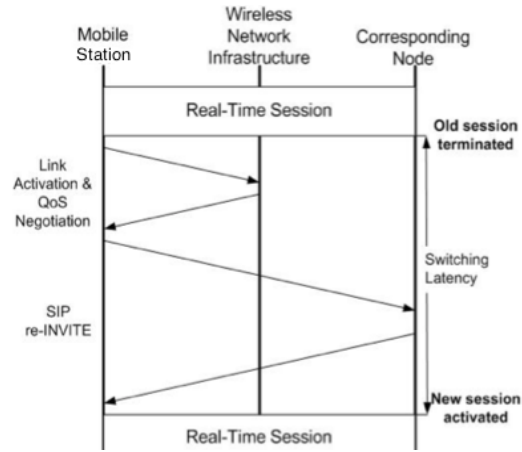

Figure 3. SIP handover for real-time session

In simulation experiments, where HO latency is considered as the time between the termination of old session and activation of new one, the HO in integrated architecture performs much better in terms of HO latency and packet loss than the case where user has subscriptions to both networks with no integration. Although the proposed solution maintains QoS provisioning between UMTS and WiMAX, it still suffers from huge packet loss during service interruption time caused by hard handover where the MS is not able to send or receive data packets.

In [4], an architecture for UMTS-WiMAX is introduced based on tight coupling interworking and using a single radio MS. The mobility between two access networks is achieved by the MIP protocol. Figure 4 describes the proposed architecture where WiMAX access network is connected to 3GPP core network via WiMAX Access Gateway (WAG) for the authentication process. Packet Data Gateway (PDG) is mainly used to route packets from or to Internet and performs FA functionality of WiMAX network. The UMTS FA functions are implemented in the GGSN. In order to enable the vertical handover between these two technologies, the HA is placed in the Internet and manages FAs of both WiMAX and UMTS.

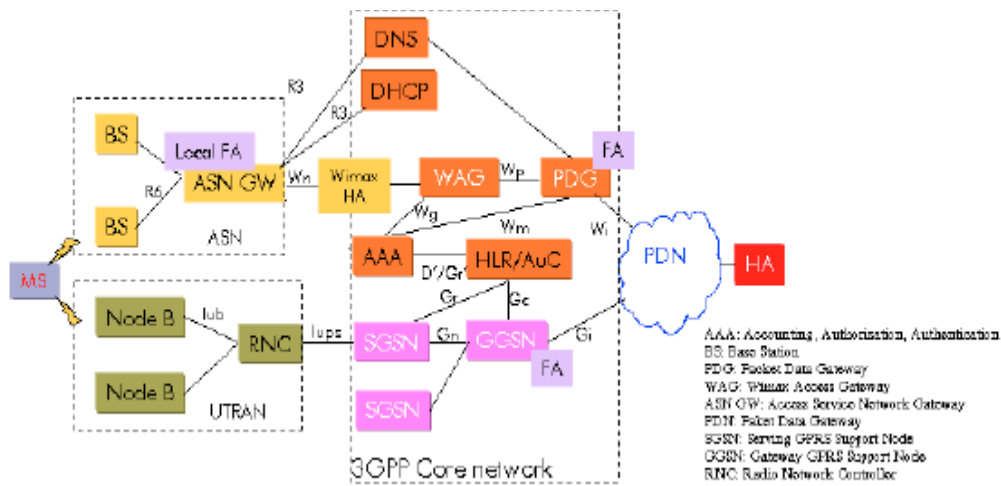

Figure 4. UMTS-WiMAX tight coupling architecture

The author proposed a VHO mechanism that allows single radio MS to prepare a new attachment in the target network before leaving the serving network. In this context, old FA notifies the HA about MS's movement so that the HA can buffer the packets and forward them to the MS as soon as the HA receives the MIP update from the MS. The specified VHO takes advantage of the total cooperation between the UMTS and WiMAX networks operators where UMTS Node B can forward WiMAX neighbouring cell information to the MS the and vice versa. The proposed tight coupling solution adopts buffering in HA to reduce packet loss and allows WiMAX traffic pass by UMTS gateway to eliminate congestion. However, using the buffering technique may minimize number of packet lost but can't solve the delay in receiving data caused by HO. Besides, making UMTS Node Bs to handle WiMax BS advertisement requires upgrading existing systems which increase cost and complexity. Finally, this solution is not suitable for real-time session since network characteristics change but the session parameters remain the same such as switching high quality session from WiMAX to UMTS where the streaming speed decreases and QoS deteriorates.

Hybrid interworking architecture of loose and tight coupling with MIP support is proposed in [18]. The architecture is depicted in Figure 5. BS of WiMAX network is connected to Interworking WiMAX (I- WiMAX) Gateway. I-WiMAX performs several functions like that of local AAA server, NAT, DHCP, DNS and handles mobility management. It is assumed that MS is capable to support simultaneous communication with UMTS and WiMAX and has prior knowledge about a WiMAX operator, which has interworking support and service agreement with its UMTS operator. Based on this information, MS scans for the operators' ID, registers with a selected BS and assigned a local IP address from the I-WiMAX gateway. Then MS sends its identity information to I-WiMAX, which performs local authentication and sends MS's identity to the corresponding 3G operator's AAA server. If the authentication fails the MS will be disconnected. Otherwise, two different scenarios for connectivity to different networks are proposed.

The first scenario is based on loose coupling where 3GPP AAA sends user profile to the Local AAA server (I-WiMAX), which performs mapping between MS's Local IP and Remote IP. At that moment, user will be able to access Internet through I-WiMAX. In the second scenario, based on tight coupling, the PDG is used only for accessing the operators IP network and there is no need for sending user's profile. It's assumed that the address assigned to MS is routable within the UMTS network. The I-WiMAX gateway would identify the user traffic, based on the destination information and route it to one of its two interfaces. Details of second scenario are presented in Figure 6.

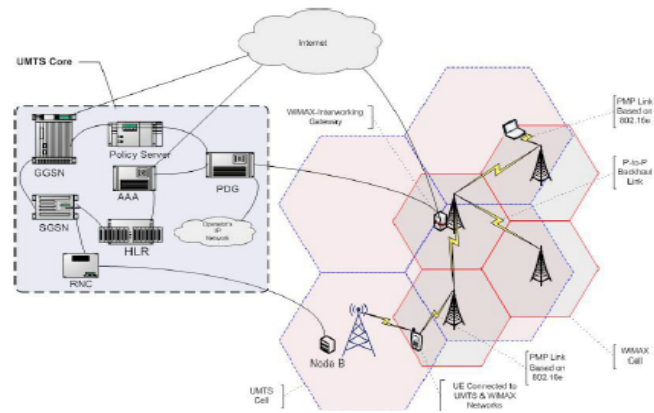

Figure 5. Hybrid interworking of UMTS-WiMAX

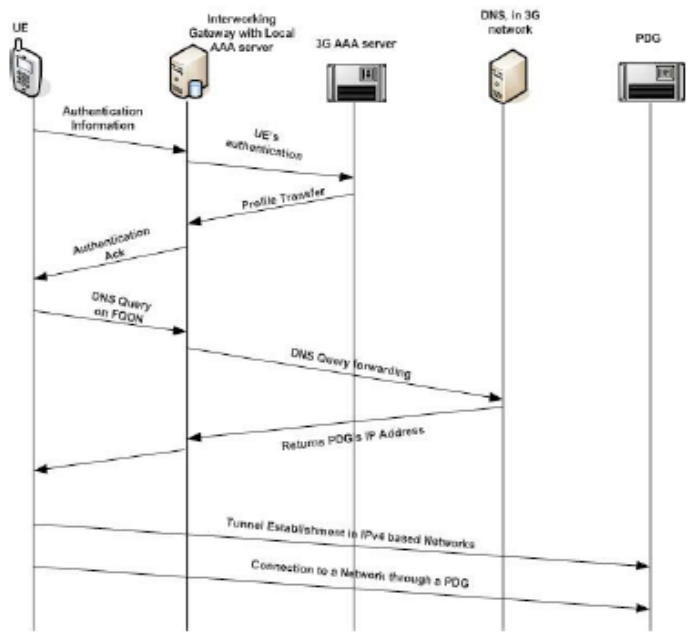

Figure 6. Tight coupling scenario

This solution takes advantage of using dual radio MS to provide seamless handover and local authentication to support a remote enforcement of operators' policies. The use of hybrid architecture with above scenarios can reduce load on UMTS core. However, the authors didn't specify a mechanism for maintaining QoS while switching access network. Besides that, having two active interfaces decreases mobile power. Also, if the authentication fails and MS is disconnected, the signalling for registration and obtaining IP address becomes unnecessary. Having more signalling can effect overall system performance and consumes MS battery.

[19] introduced an interesting cross-layer optimization for MIPv6 and FMIPv6 handover between Mobile WiMAX and 3G networks. The proposed cross-layer solution reorders, palatalizes and combines L2 and L3 signalling messages to get shorter handover latency and lower signalling load. To achieve that with MIPv6, L3 information of target FA is delivered through link layer (L2) message and RtSol/RtAdv messages are omitted. In addition, reordering of L2 and L3 messages is performed in WiMAX network. The optimized MIPv6 handover to WiMAX and to UMTS is shown in Figure 7 and Figure 8.

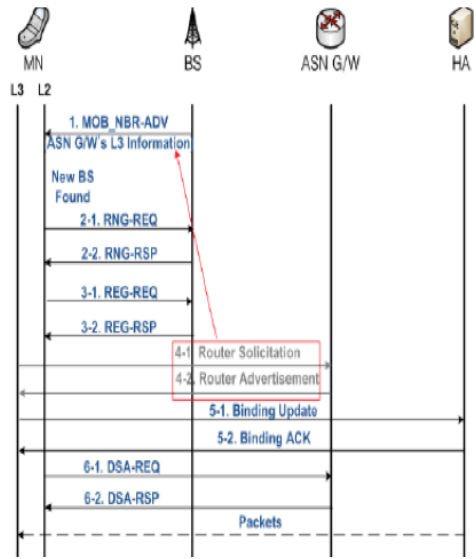

Figure 7. Cross layer MIPv6 to WiMAX

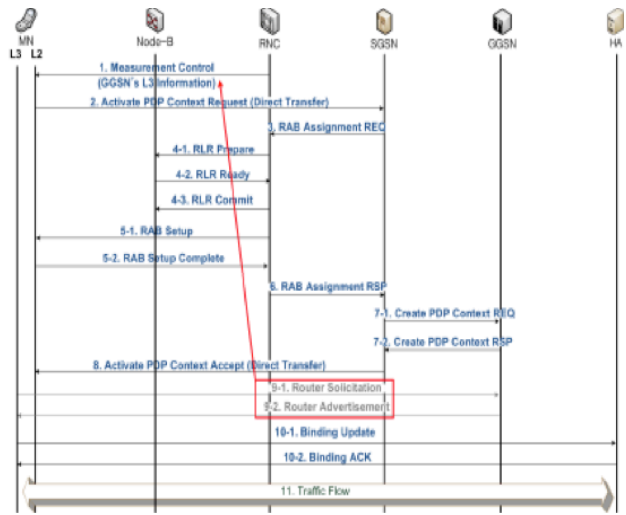

Figure 8. Cross layer MIPv6 to UMTS

Similar to MIPv6, L3 information of target network is delivered in L2 message and the RotSolPr/PrRtAdv are omitted in FMIPv6. In addition, FNA message, which indicates all the procedures are done and buffered packets can be forwarded to MS, can be included in L2 signalling message too. In predictive mode, FNA (L3 message) is omitted and L2 message will contain Link Layer Address (LLA) of MS. In reactive mode FNA plays a role to carry NCoA of MS. The DSA-REQ and RAP-Setup Complete L2 messages of WiMAX and UMTS respectively, will carry NCoA of MS. The cross layer predictive and reactive modes to Mobile WiMAX are illustrated in Figure 9 and Figure 10, respectively.

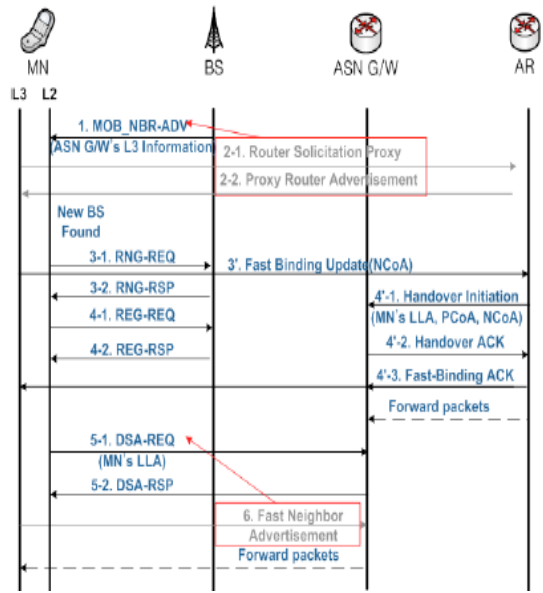

Figure 9. Cross layer FMIPv6 predictive mode

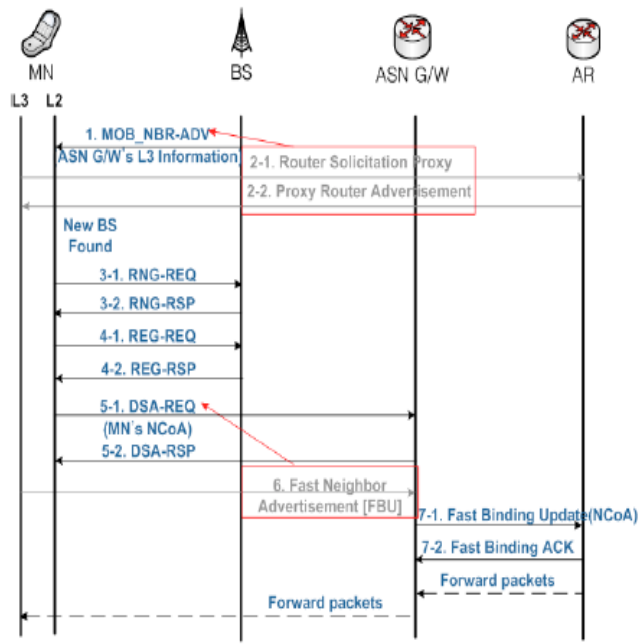

Figure 10. Cross layer FMIPv6 reactive mode

As can be provisioned the number of signalling messages is reduced. Simulation results show that the cross layer handover performs much better compared to basic MIPv6 and FMIPv6 in terms of latency, packet delay and loss. However, this requires modifications on WiMAX and 3G standards because of the added functionality in the standard L2 signalling messages [19].

[17] presented a novel inter-RAT (Radio Access Technology) VHO scheme between UMTS and Mobile WiMAX based on very tight coupling architecture. The proposed model introduced a common interworking sublayer (IW sublayer) at layer 2 on both RNC and MS. The IW takes the role of LLC sublayer of conventional cellular networks, such as retransmission and handover support, determines a suitable target network, create primitive between the IW and the

UMTS network or between the IW and the WiMAX network and performs SR ARQ (Selective Repeat ARQ) mechanism. The architecture is depicted in Figure 11.

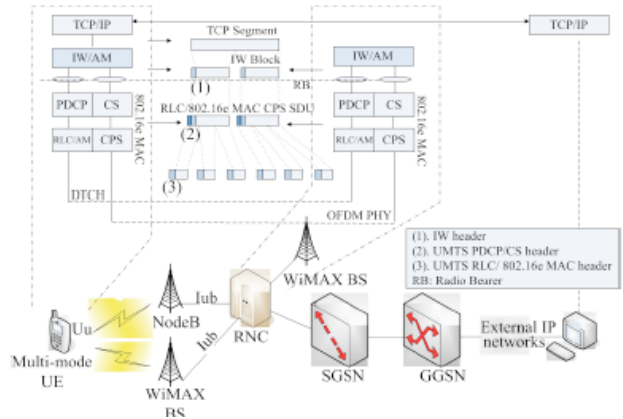

Figure 11. Interworking architecture of Mobile WiMax and UMTS with IW

In the proposed solution, packet is segmented into smaller IW blocks, each of which is assigned a sequence number. This new IW sub-header is used for block loss detection and block re-sequencing in the receiver to guarantee in-sequence delivery. IW queue transmitted packets for retransmission if needed. When handover is triggered, IW communicates with target network to prepare for the handover, starts buffering the incoming packets and sends handover command to MS via source network. The MS performs hard handover by releasing old link and then start the radio link setup with the target network. After the connection is completed, notification from target BS or RNC is sent to IW to restart transmitting of packets. Simulation shows that IW sublayer can eliminate packet loss for hard handover but suffers from late packet delivery. For signalling flow of handover from UMTS to WiMAX refer to Figure 12.

The authors extended their work in [20] to support soft and make-before-break handover by using mutilhoming technique. Soft handover is applied for UMTS to WiMAX case and is achieved using bi-casting technique. In this regard, when IW sublayer in RNC receives notification that MS is connected to WiMAX, it duplicates packets destined for MS and forwards them through UMTS and WiMAX links simultaneously. These duplicate packets are merged at IW sublayer on the MS. The IW sublayer continues bi-casting until it receives link going down trigger from UMTS. This trigger makes IW stop bi-casting and soft handover is completed. Regarding handover from WiMAX to UMTS, the make-before-break handover is used instead, because the negative effect, caused by different bandwidth and transmission delay between two wireless links, becomes significant to limit the soft handover performance gain. The only difference between this approach and the one described in [17] is that IW continues forwarding packet through WiMAX interface until it receives notification from UMTS that the MS has successfully MS finishes UMTS radio link setup. Upon receiving this notification, the IW only adjusts its local retransmission windows size and continues the data block forwarding through UMTS link.

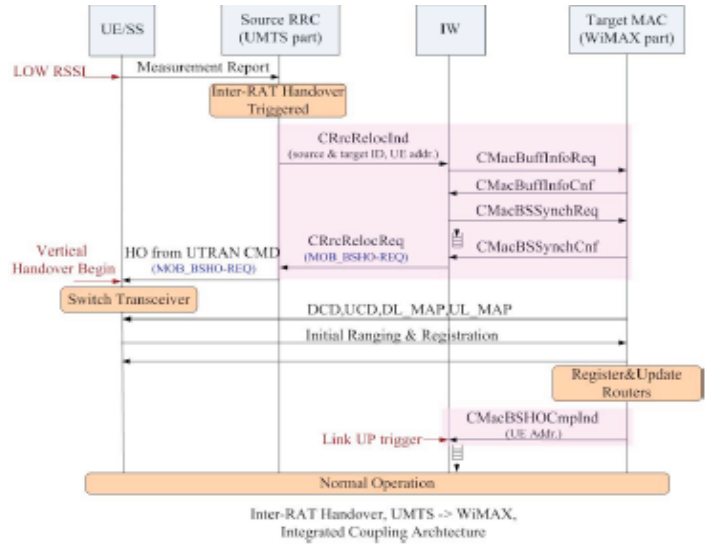

Figure 12. UMTS to WiMAX L2 handover

The proposed mechanism solves the problem of late packet delivery as well as the need for buffering and can provide seamless VHO. However, it suffers from the problem of network load and resource wastage. Not to mention high cost and complexity required for the modification of the existing systems. The iner-RAT or link layer solutions are expected to provide the best vertical handover performance, but usually not preferable, due to technology-specific modifications that require standardization and the usual restriction on the single operator. Table 1 describes the advantage and disadvantage for each of loose and tight coupling solutions.

Table 1. Loose and tight coupling solutions.

| Solution | Maintain QoS | Less Packet Loss | Features | Drawbacks |
|---|---|---|---|---|
| [16] | ✓ | | • Offers a suitable QoS at a given network<br>• Local authentication | • Packet loss<br>• High HO latency |
| [4] | | ✓ | • Low HO delay | • Packet delay<br>• Requires buffering |
| [18] | | ✓ | • Local authentication | • Power consumption<br>• Requires upgrading of UMTS CN |
| [19] | | ✓ | • Minimize signaling<br>• Low HO delay | • Modifications are required |
| [17] | | ✓ | • No packet loss<br>• Low HO delay | • Modifications are required<br>• Packet delay |
| [20] | | ✓ | • No packet loss<br>• No packet delay | • Bi-casting wastes resources<br>• Scalability problem |

## 4.2. 3GPP IMS-based solutions

The 3GPP IMS comes as a promising solution for integrating UMTS and WiMAX since the system offers the needed interworking environment independent of any access technology. To resolve the problems of session continuity and keep IP address of MS unchanged during VHO between WiMAX and UMTS, [2] and [3] proposed methods to use 3GPP IMS and MIP for session continuity and maintaining IP address respectively.

[3] proposed a novel architecture to connect 3G (3GPP or 3GPP2) and Broadband Wireless Access (BWA) system, such as WiMAX or WLAN networks, with 3GPP IMS as an arbitrator for interworking. The MIPv4 is used to handle mobility where the FAs are deployed in GGSN, CNS and PDSN of each of the ANs. Figure 13 shows the proposed architecture where each network is connected to the all-IP CN and a local P-CSCF is deployed in each of them. The HA of MIP and remaining elements of the IMS are located at the home network of the MS. The proposed VHO is based on make-before-break handover mechanism. The VHO is triggered when MS enters the WiMAX coverage and while data flow via UMTS is still active, MS performs standard WiMAX link layer access registration and the MIP registration followed by SIP Re-INVITE message for session re-establishment through WiMAX interface. Then resources are reserved in WiMAX and the new data flow is initiated via WiMAX interface. Detailed session establishment and handover are shown in Figure 14.

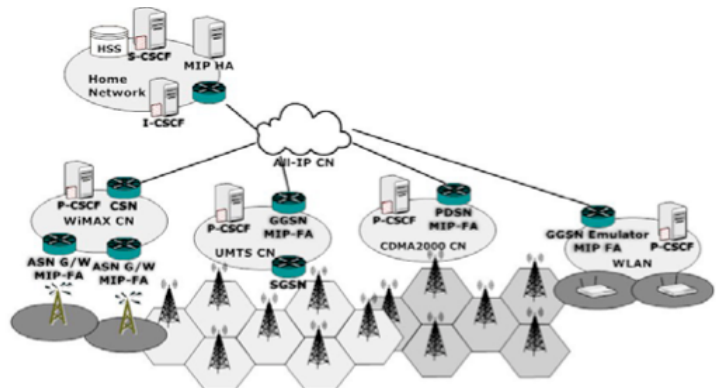

Figure 13. 3GPP, 3GPP2 and BWA integration architecture

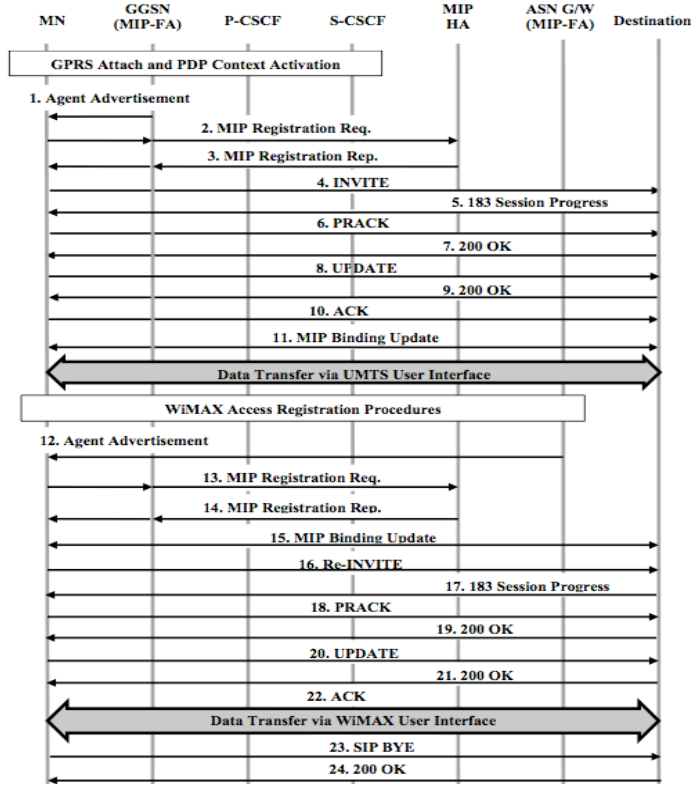

Figure 14. MIP-SIP signalling within IMS

Simulation results indicate that the handover delay is always smaller when MIP is used in contrast to a pure SIP mechanism, which incurs additional IMS based processing latencies. The use of SIP and make-before-break handover can maintain QoS of session and minimize packet loss, respectively. However, using dual radio MS increases battery consumption and using MIP in conjunction with SIP increases signalling load.

[2] has the same architectural elements as [3] with use for application server (AS) of IMS as the CN of MIP. The novelty of this approach resides in distributing server bandwidth among the networks according to the speed supported. Also, the server channel frequency is kept less than BS channel frequency so that there remains no in-transit packet in queue and to eliminate packet loss in the radio path during handover. The proposed VHO algorithm is illustrated in Figure 15. Results indicate this solution is very effective in reducing handover latency and eliminating packet loss in radio path. The solution still suffer from the problem of signalling load and battery consumption.

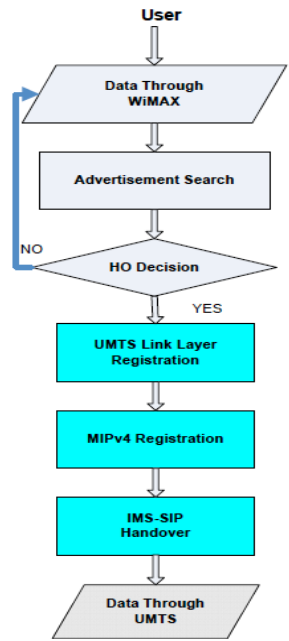

Figure 15. WiMAX to UMTS handover algorithm

[21] proposed a tight coupling interworking model for UMTS, WiMAX and WLAN in 3GPP IMS architecture. The architecture in Figure 16 incorporates a UMTS Core Network where SGSN, GGSN and AAA Server reside. WiMAX and WLAN interconnected with UMTS CN through specific functional entities and IMS is in charge of sessions' control. Users can access UMTS CS based services using any of the integrated access technologies, since they are authenticated in the AAA Server and registered in the IMS. Each AN is connected to a distinct PDG and to a distinct WAG (Wireless Access Gateway). The same P-CSCF serves ANs' PDGs and WAGs, whilst different S-CSCF and I-CSCF servers could be allocated for each AN. The MSs are identified by multiple IP addresses and have the ability to choose the AN. A prediction handover procedure is used to eliminate the service disruption time. Also, during handover, information concerning the active calls is sent to the new P-CSCF from old P-CSCF serving the user to reduce amount of time needed for the handover. The handover starts with the MS attaching to UMTS core and activating a PDP context followed by the IMS reregistration including context transfer mechanism. Then MS re-invites the CN to the session. At this point, the new P-CSCF has to authorize the QoS class of the session based on the service classes' correlation. The re-invite request is forwarded to CN before the data flow can be re-established. Handover signalling is shown in Figure 17.

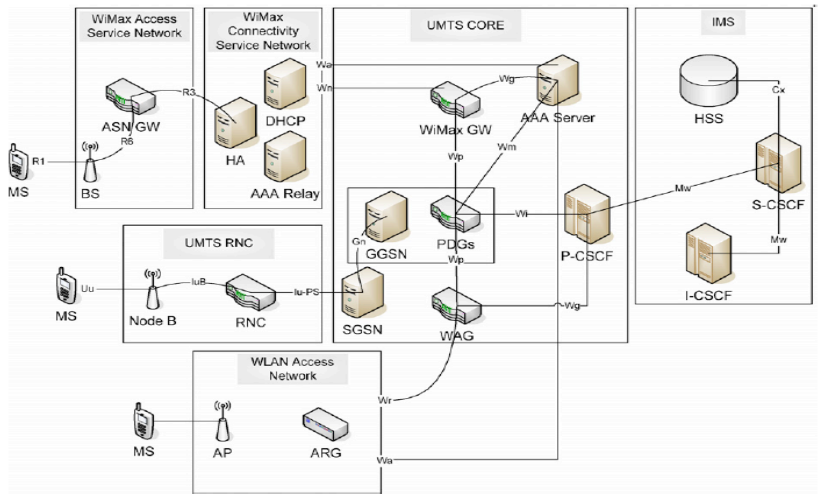

Figure 16. IMS interworking of UMTS-WiMAX-WLAN

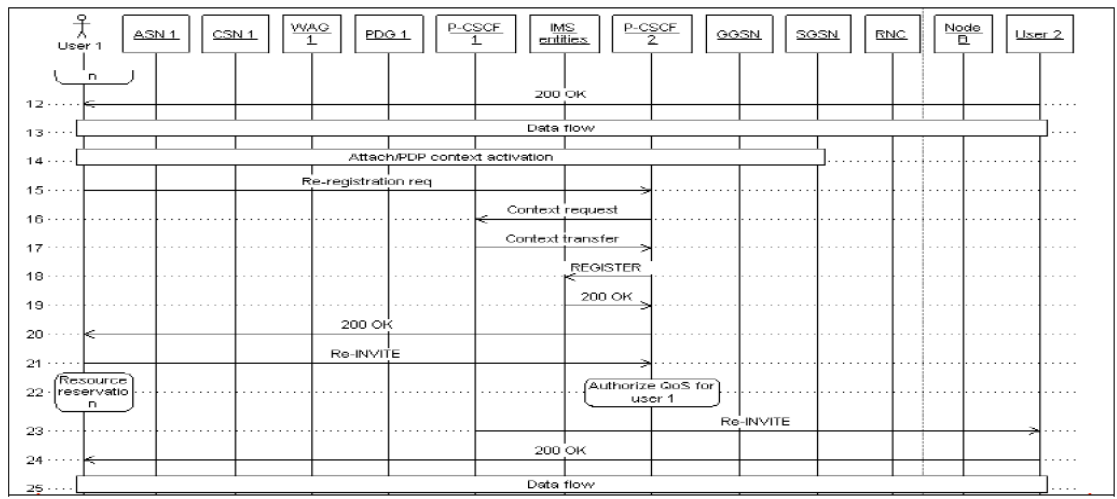

Figure 17. Handover signalling from WiMAX-UMTS

This solution can provide faster handover with less signalling compared to MIP-SIP handover. However, it has a scalability problem as the number of ANs increase the load will increase on P-CSCF. Another problem is the modification required on existing systems to support tight coupling. The features and drawbacks of IMS-based solutions are depicted in Table 2.

Table 2. IMS-based solution.

| Solution | Maintain QoS | Less Packet Loss | Features | Drawbacks |
|---|---|---|---|---|
| [3] | ✓ | ✓ | • Maintain QoS and MS IP<br>• Low HO delay | • Power consumption<br>• High signalling |
| [2] | ✓ | ✓ | • No packet loss in radio bath | • Power consumption<br>• High signalling |
| [21] | ✓ | ✓ | • Access to UMTS CS<br>• Fast HO | • Scalability problem<br>• Modifications |

## 4.3. EPC based solutions

Recently some researchers have been working on integrating Mobile WiMAX and 3GPP networks based on Evolved Packet Core (EPC).

[22] proposed an interworking architecture for integrating Mobile WiMAX with 3GPP networks, specifically UMTS, based on the specified 3GPP EPC and PMIPv6, with an optimized VHO mechanism to support seamless mobility between the two networks for single radio MS. A new functional element called Forward Authentication Function (FAF) has been introduced in EPC. The FAF is a BS-level entity that is located in the target access. It supports the authentication of the MS before the execution of VHO through the IP tunnel. The FAF performs the functionalities of WiMAX BS when the MS is moving toward a WiMAX cell, or it may also perform as a UMTS RNC. In VHO process it is assumed that both networks support PMIPv6, so a PMIPv6 tunnel has been established between the source network and the P-GW. When MS discovers ANDSF in the serving EPC, it establishes a secure connection with it and receives information about neighbouring networks as well as intersystem mobility policies. The MS then measures the discovered neighbour cells and if at least one provides adequate signal strength, it initiates preregistration procedure.

Handover from WiMAX to UMTS has been discussed in this work. During preregistration, a secured tunnel is established between MS and FAF, which receives an Attach Request message and forward it to SGSN over a common Iu-ps interface. This triggers the normal UMTS authentication procedure over the MS-FAF IP tunnel. If the authentication is successful, the SGSN accepts the attach request by sending an Attach Accept message and updates the MS location in the HSS. At this point preregistration process is over and neither the SGSN, nor any other network element in the core 3GPP network knows that the MS has attached while still being on WiMAX access. If there is a need for handover, MS transmits a Handover Request message to the FAF including the target UTRAN cell to which it wants to handover and FAF prepares the appropriate radio resources there then responds with a Handover Command message that includes information about the target UTRAN cell. The MS then disconnect from WiMAX and perform normal handover execution according to the UMTS specifications. The PMIPv6 tunnel in P-GW is relocated from the WiMAX ASN-GW to the S-GW during the PDP Context creation and S-GW updates this tunnel by sending a new Proxy MIP Binding Update, which updates the mobility information in the P-GW. The MS maintains connectivity with the same P-GW and maintains the same IP address by including special information in the PDP Context Activation Request message that helps the S-GW select the same P-GW already allocated to this MS. Figure 18 shows signalling flow between networks components during handover.

The proposed solution introduced a mechanism to provide make-before-break handover for the single radio MS. This mechanism reduces handover latency and packet loss for single radio MS with minimum change to existing systems by introducing FAF to handle preregistration and resource allocation in the target network. However, one of main problems of this solution is that the MS does not make any effort to inform the source network of the VHO operation and leaves the source network without proper disconnection procedures. This results in loss of buffered data packets and keeps the resources unreleased in the source network after the MS disconnects.

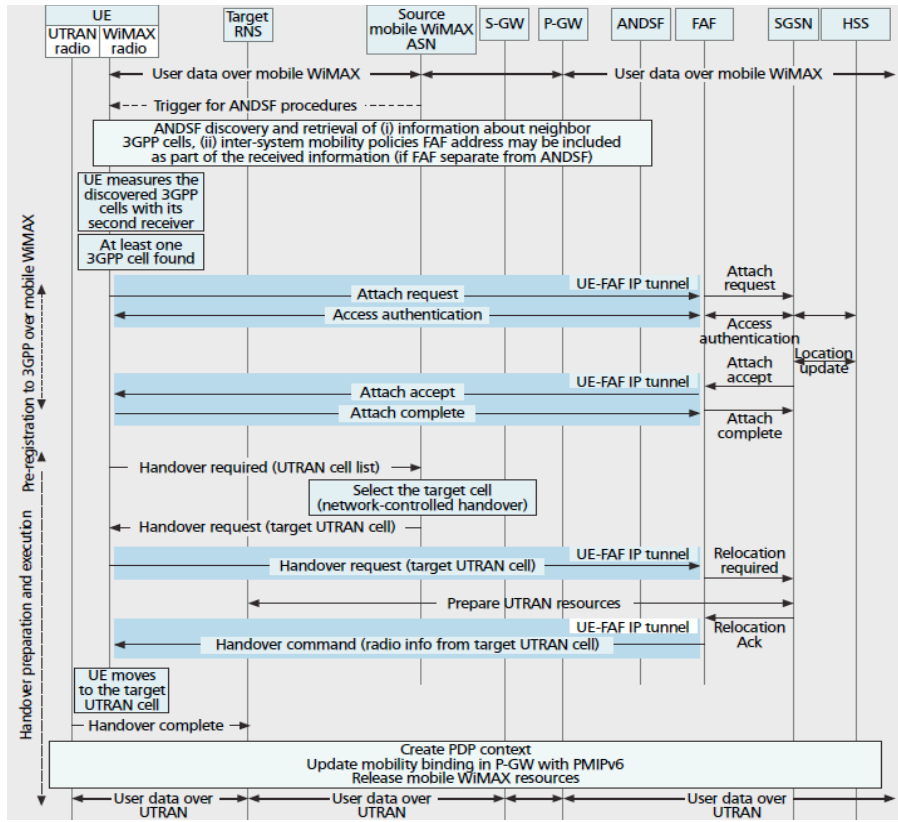

Figure 18. WiMAX to UMTS handover in EPC with FAF

In [23], a Data Forwarding Function (DFF) is added to the source network for resolving the problem of packet loss and abnormal source network disconnection. The DFF is BS-level like FAF and can emulate network BS where the MS is located in. It handles the data forwarding and deregistration procedures. The MS communicates with DFF via IP tunnel and IP address of the FAF is sent to DFF via that tunnel. The DFF establishes the IP connection to the designated FAF and performs WiMAX BS or intra-SGSN handover to retrieve the user data buffered in the source access. Then it forwards packets to the FAF and performs deregistration process in the source access after the MS has moved to the target access and data forwarding is completed. The authors discussed VHO from mobile WiMAX to 3GPP UTRAN, where DFF emulates the mobile WiMAX BS and the FAF performs as an UTRAN RNC. The proposed architecture is illustrated in Figure 19 and the handover signalling is shown in Figure 20.

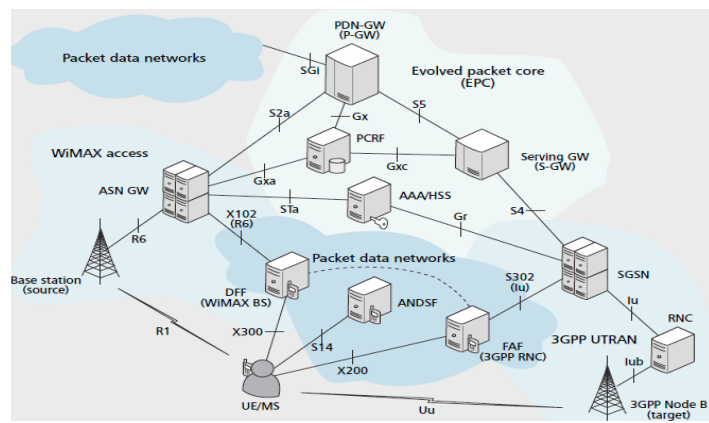

Figure 19. Interworking architecture with FAF and DFF

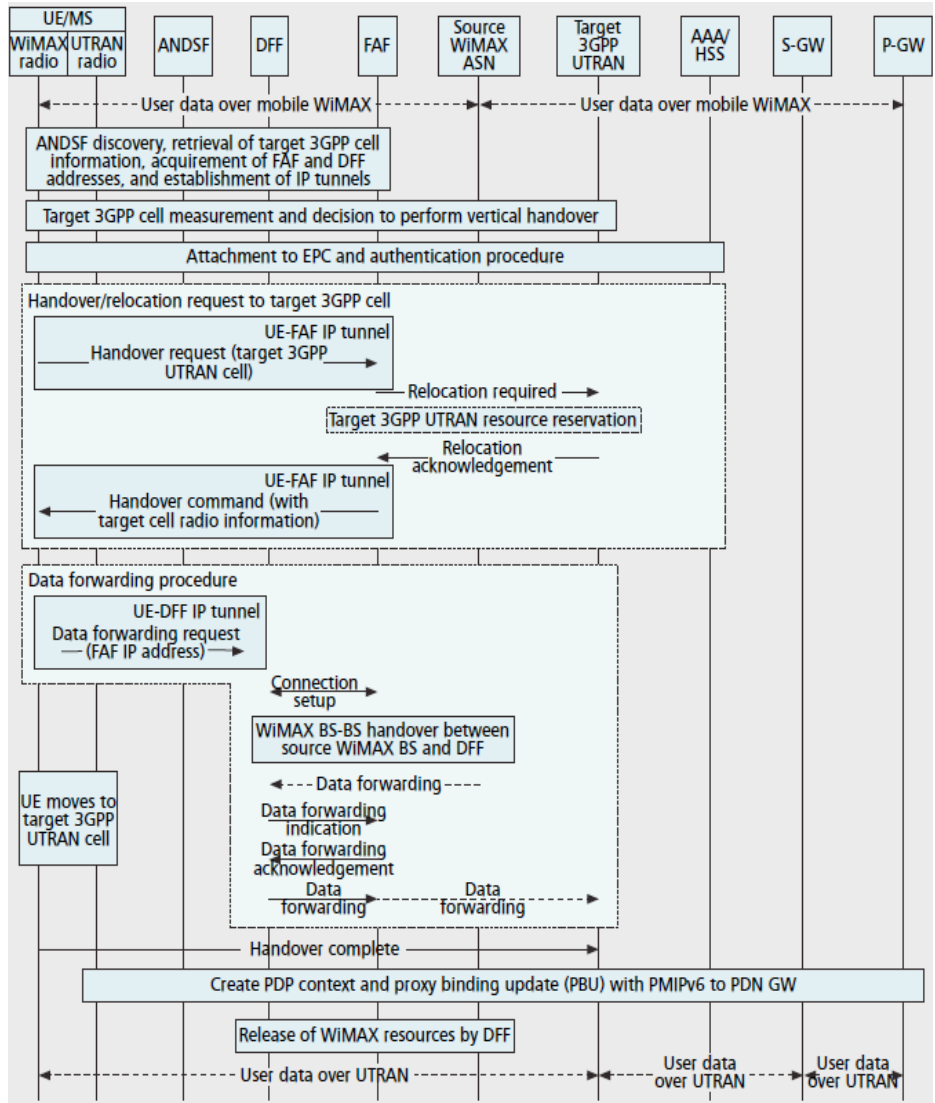

Figure 20. Signalling flow for VHO from WiMAX to UMTS with FAF and DFF

The simulation results of heavily multiplexed real-time video traffic using DFF show potential improvements in packet loss compared to [22]. However, communication between additional entities and MS increases signalling and additional information are required to be stored in ANDSF. The features and drawbacks of the two solutions are depicted in Table 3.

Table 4. EPC-based solutions.

| Solution | Maintain QoS | Less Packet Loss | Features | Drawbacks |
|---|---|---|---|---|
| [22] | ✓ | ✓ | • Minimize HO delay for single radio MS <br> • Resource allocation | • Unreleased resource at source network <br> • On air packets are lost |
| [23] | ✓ | ✓ | • Packet forwarding <br> • Resource released | • More signalling <br> • Additional information are required to be stored in ANDSF |

## 5. COMPARISON OF UMTS-WiMAX HANDOVER

In this section, a brief comparison of different VHO solutions is summarized in Table 4. The comparison is based on a set of criteria including: architecture, mobility management protocol, type of handover and whether or not new entities are required. The table also includes main network elements that are involved in handover for each solution.

Table 4. Comparison of UMTS-WiMAX handover.

| Solution | Architecture | Mobility Management Protocol | Handover Type | New Entities | Signalling Flow |
|---|---|---|---|---|---|
| **[16]** | Loose coupling HA, Home AAA, SIP server, FAs MS Architecture | MIP for non real SIP for real time applications | Hard | | FA, HA, local AAA, home AAA Or Re-invite to CN |
| [4] | Tight coupling HA, FAs | MIP | Make-before-break | | FA, HA |
| **[18]** | Loose-Tight coupling | MIP | Make-before-break | | FA, HA, I-WiMAX, home AAA |
| **[19]** | Loose coupling HA, FAs | Cross layer MIPv6 Cross layer FMIPv6 | Hard | | FA, HA |
| **[17]** | Very Tight Interworking sublayer at L2 on RNC and MS | UTRAN mobility mechanism /Hard HO | Hard | ✓ | RRC, IW, MAC |
| **[20]** | Tight- Very Tight Interworking sublayer at L2 on RNC and MS | UTRAN mobility mechanism | UMTS->WiMAX Soft WiMAX->UMTS Make-before-break | ✓ | RRC, IW, MAC MAC, IW, RRC |
| **[3]** | 3GPP IMS HA, FAs | MIP-SIP | Make-before-break | | FA, HA, P-CSCF, S-CSCF |
| **[2]** | 3GPP IMS HA, FAs, AS | MIP-SIP | Make-before-break | | FA, HA, P-CSCF, S-CSCF |
| **[21]** | 3GPP IMS Tight Coupling | SIP | Make-before-break | ✓ | P-CSCF2, P-CSCF1 |
| **[22]** | EPC FAF | PMIPv6 | Make-before-break | ✓ | ANDSF, FAF |
| **[23]** | EPC FAF, DFF | PMIPv6 | Make-before-break | ✓ | ANDSF, FAF, DFF |

## 3. CONCLUSIONS

In this paper, various interworking architectures and handover scenarios of UMTS and WiMAX have been discussed. From an architectural point of view, proposed solutions are classified into three categories: loose and tight coupling, IMS-based and EPC-based solutions. While the first category is focusing on seamless mobility or QoS provisioning, the IMS and EPC are taking the two problems into consideration. However, as can be provisioned there is always a trade-off between providing seamless mobility with QoS provisioning and signalling overhead which can effect the overall system performance and consume MS battery. Thus, providing user with seamless mobility with QoS supports remains an open issue. Future work should focus on providing a mechanism to reduce signalling overhead with minimum change to the existing systems and protocols.

ACKNOWLEDGEMENTS

This paper contains the results and findings of a research project that is funded by King Abdulaziz City for Science and Technology (KACST) Grant No: A-S-11-0463.

**Authors**

Nada Alamri received her degree in Computer Science from Um Alqura University, Saudi Arabia, 2007. Since 2009 she has been at King Abdulaziz University, where she is now a master student.

Nadine akkari has an over 8 years of experience in the fields of network engineering, with focus on new emerging technology and all IP network Design in terms of Quality of Services, mobility management and seamless handovers. She studied Computer Engineering at University of Balamand, Lebanon, 1999. She got her Masters in telecommunications networks from Saint-Josepf University, Lebanon in 2001and her PhD in Mobility and QoS Management in Next Generation Networks in 2006 from National Superior School of telecommunications (ENST), Paris, France. Currently she is assistant professor in the faculty of Computing and Information Technology in King Abdulaziz University, Jeddah, Saudi Arabia.